\documentclass[fleqn,12pt]{article}
\usepackage{latexsym,amssymb,cite}

\def\noi{\noindent}

\newcommand{\Title}[1]{\noi {{\LARGE\bf #1}} \\}
\newcommand{\Author}[2]{\noi{\large\bf #1}\\[2ex]\noindent{\it #2}\\}

\newcommand{\Abstract}[1]{\vskip 2mm \begin{center}
        \parbox{16.4cm}{\small\noi #1} \end{center}\medskip}

\newcommand{\foom}[1]{\protect\footnotemark[#1]}

\newcommand{\email}[2]{\footnotetext[#1]{e-mail: #2}
		\addtocounter{footnote}{1}}

\newcommand{\Theorem}[2]{\medskip\noi {\bf #1. \ }{\sl #2}\medskip}

\newcommand{\sect}[1]{Sec.\,#1}
\def\nq{\hspace*{-1em}}
\def\nqq{\hspace*{-2em}}
\def\nhq{\hspace*{-0.5em}}

\def\cm{\hspace*{1cm}}

\def\Jl#1#2{#1 {\bf #2},\ }

\def\ApJ#1 {\Jl{Astroph. J.}{#1}}
\def\CQG#1 {\Jl{Class. Quantum Grav.}{#1}}
\def\DAN#1 {\Jl{Dokl. AN SSSR}{#1}}
\def\GC#1 {\Jl{Grav. Cosmol.}{#1}}
\def\GRG#1 {\Jl{Gen. Rel. Grav.}{#1}}
\def\JETF#1 {\Jl{Zh. Eksp. Teor. Fiz.}{#1}}
\def\JETP#1 {\Jl{Sov. Phys. JETP}{#1}}
\def\JHEP#1 {\Jl{JHEP}{#1}}
\def\JMP#1 {\Jl{J. Math. Phys.}{#1}}
\def\NPB#1 {\Jl{Nucl. Phys.}{B\ #1}}
\def\NP#1 {\Jl{Nucl. Phys.}{#1}}
\def\PLA#1 {\Jl{Phys. Lett.}{#1A}}
\def\PLB#1 {\Jl{Phys. Lett.}{#1B}}
\def\PRD#1 {\Jl{Phys. Rev.}{D\ #1}}
\def\PRL#1 {\Jl{Phys. Rev. Lett.}{#1}}

\def\al{&\nhq}
\def\lal{&&\nqq {}}
\def\eq{Eq.\,}
\def\eqs{Eqs.\,}
\def\beq{\begin{equation}}
\def\eeq{\end{equation}}
\def\bear{\begin{eqnarray}}
\def\bearr{\begin{eqnarray} \lal}
\def\ear{\end{eqnarray}}
\def\earn{\nonumber \end{eqnarray}}
\def\nn{\nonumber\\ {}}

\def\yyy{\\[5pt] \lal }
\def\eql{\al =\al}

\def\dst{\displaystyle}
\def\tst{\textstyle}
\def\fracd#1#2{{\dst\frac{#1}{#2}}}
\def\fract#1#2{{\tst\frac{#1}{#2}}}
\def\Half{{\fracd{1}{2}}}
\def\half{{\fract{1}{2}}}
\def\thd{{\fract{1}{3}}}

\def\e{{\,\rm e}}
\def\d{\partial}

\def\sign{\mathop{\rm sign}\nolimits}
\def\diag{\mathop{\rm diag}\nolimits}

\def\const{{\rm const}}
\def\eps{\varepsilon}

\def\then{\ \Rightarrow\ }

\newcommand{\vars}[1]{\left\{\begin{array}{ll}#1\end{array}\right.}
\def\R{{\mathbb R}}
\def\tT{{\widetilde T}}
\def\rr{{\overline r}}

\def\kappa{\varkappa}
\def\opsi{\overline{\psi}}
\def\spn{{\rm sp}}
\def\tgamma{\tilde{\gamma}}

\def\mn{_{\mu\nu}}
\def\MN{^{\mu\nu}}
\def\mN{_\mu^\nu}

\def\GR{general relativity}
\def\sph{spherically symmetric}
\def\ssph{static, spherically symmetric}
\def\cy{cylindrical}
\def\cyl{cylindrically symmetric}
\def\scyl{static, cylindrically symmetric}

\def\bh{black hole}

\def\wh{wormhole}
\def\whs{wormholes}
\def\asflat{asymptotically flat}

\begin{document}

\Title {\bf Cylindrical wormholes}

\bigskip

\Author{Kirill A. Bronnikov\foom 1}
{\small Center for Gravitation and Fundamental Metrology, VNIIMS, 46
   Ozyornaya St., Moscow 119361, Russia;\\ Institute of Gravitation and
   Cosmology, PFUR, 6 Miklukho-Maklaya St., Moscow 117198, Russia}

\Author{Jos\'e P. S. Lemos\foom 2}
{\small Centro Multidisciplinar de {}Astrof\'{\i}sica - CENTRA,
Departamento de F\'{\i}sica, Instituto Superior T\'ecnico - IST,
Universidade T\'ecnica de Lisboa - UTL, Avenida Rovisco Pais 1, 1049-001
Lisboa, Portugal}

\bigskip

\Abstract
  {It is shown that the existence of \scyl\ \whs\ does not require violation
  of the weak or null energy conditions near the throat, and \cyl\ \wh\
  geometries can appear with less exotic sources than \whs\ whose throats
  have a spherical topology. Examples of exact \wh\ solutions are given with
  scalar, spinor and electromagnetic fields as sources, and these fields
  are not necessarily phantom. In particular, there are \wh\ solutions for a
  massless, minimally coupled scalar field in the presence of a negative
  cosmological constant, and for an azimuthal Maxwell electromagnetic field.
  All these solutions are not \asflat. A no-go theorem is proved, according
  to which a flat (or string) asymptotic behavior on both sides of a \cy\
  \wh\ throat is impossible if the energy density of matter is everywhere
  nonnegative.
\medskip

}

%%%%%%%%%%%%%%%%%%%%%%%%%%%%%%%%%%
\email 1 {kb20@yandex.ru}
\email 2 {lemos@fisica.ist.utl.pt}

\section{Introduction}

  Lorentzian traversable wormholes as smooth bridges between different
  universes, or smooth shortcuts between remote parts of a single universe,
  have been widely discussed from different theoretical standpoints for many
  years, see \cite{vis-book,HV97,LL-03,Sush05} for reviews. It is well known
  that a \wh\ geometry can only appear as a solution to the Einstein
  equations if the stress-energy tensor (SET) of matter violates the null
  energy condition (NEC) at least in a neighborhood of the \wh\ throat
  \cite{HV97}. This conclusion, however, has been proved under the
  assumption that the throat is a compact 2D surface, having a finite
  (minimum) area (at least in the static case, since for dynamic \whs\ it has
  proved to be necessary to generalize the notion of a throat \cite{HV98};
  see, however \cite{HAY98} for other definitions of a throat). In other
  words, it was implied that, as seen from outside, a \wh\ entrance is a
  local object like a star or a \bh.

  But, in addition to such objects, the Universe may contain structures
  which are infinitely extended along a certain direction, like cosmic
  strings. And, while starlike structures are, in the simplest case,
  described in the framework of spherical symmetry, the simplest stringlike
  configurations are \cyl. Their possible \wh\ properties will be the
  subject of the present paper. For simplicity, we here consider only static
  configurations. It should be stressed that we will deal with genuine
  cylindrical symmetry, unlike Kuhfittig \cite{kuhf} who considered
  cylindrical configurations of finite length, which are actually axially
  symmetric due to $z$ dependence. A special case of \whs\ related to
  the present setting (with a cosmic string metric and thin shells of
  negative density) has been considered in \cite {sim04-06}.

  For \ssph\ space-times with the metric
\beq                                                         \label{ds-sph}
    ds^2 = A(u) dt^2 - B(u) du^2 - r^2(u)
    				(d\theta^2 + \sin^2 \theta d\phi^2)
\eeq
  (where $u$ is an arbitrary admissible spherical radial coordinate), we say
  that there is a \wh\ geometry if at some $u=u_0$ the function $r(u)$ has a
  regular minimum $r(u_0) > 0$ (which is then called a throat) and, on both
  sides of this minimum, $r(u)$ grows to much larger values than $r(u_0)$.
  It is supposed that, at least in some range of $u$ containing $u_0$, the
  functions $A(u)$ and $B(u)$ are also smooth, finite and positive, which
  guarantees regularity and absence of horizons.\footnote
	{We thus do not restrict ourselves to \asflat\ \whs\ and, moreover,
	admit that a horizon may occur somewhere far from the throat, as it
	happens, e.g., if a \wh\ is asymptotically de Sitter due to a
	small cosmological constant.}

  Likewise, consider \scyl\ space-times with the general metric taken in
  the form
\beq                                                         \label{ds-cy}
    ds^2 = \e^{2\gamma(u)} dt^2 - \e^{2\alpha(u)} du^2 -
		\e^{2\xi(u)} dz^2 - \e^{2\beta(u)} d\phi^2
\eeq
  where $u$ is an arbitrary admissible \cy\ radial coordinate, $z\in \R$ is
  the longitudinal coordinate, and $\phi \in [0,\ 2\pi]$ is the angular one.
  The main global features of such space-times are defined in terms of the
  behavior of the circular radius $r(u) = \e^{\beta(u)}$: namely, a spatial
  asymptotic (if any) corresponds to $r(u) \to \infty$, and a symmetry axis
  (if any) is defined by vanishing $r(u)$, which means that the coordinate
  circles shrink to points. It therefore seems reasonable to accept the
  following definition of a \cy\ \wh:

\Theorem{Definition 1}
  {We say that the metric (\ref{ds-cy}) describes a \wh\ geometry if
  the circular radius $r(u)$ has a minimum $r(u_0) > 0$ at some $u=u_0$,
  if, on both sides of this minimum, $r(u)$ grows to much larger values than
  $r(u_0)$, and, in some range of $u$ containing $u_0$, all four metric
  functions in (\ref{ds-cy}) are smooth and finite {\rm (which guarantees
  regularity and absence of horizons)}.

  The cylinder $u=u_0$ is then called a throat.}

  The notion of a \wh\ is, as in other similar cases, not rigorous
  because of the words ``much larger'', but the notion of a throat as
  a minimum of $r(u)$ is exact. Asymptotically regular \whs, to be discussed
  below, are also defined exactly.

  We will give some examples of \wh\ solutions whose sources are scalar,
  nonlinear spinor and electromagnetic fields. It is important that these
  fields need not be phantom, and, in particular, there are Einstein-Maxwell
  \wh\ solutions with azimuthal electric or magnetic fields.

  It is also possible to define a \cy\ \wh\ by analogy with \sph\ or
  other starlike configurations, using, instead of $r(u)$, the area function
  $a(u) = \e^{\beta + \xi}$ of 2D \cy\ surfaces $t=\const,\ u=\const$. (Their
  area is certainly infinite but becomes finite if we identify some points on
  the $z$ axis, thus converting \cy\ symmetry to toroidal.) We still believe
  that Definition 1 is more appropriate for \cy\ symmetry, although it is
  useful to compare the properties of different notions of a throat.

  The paper is organized as follows. In \sect 2\ we present necessary
  conditions for the existence of a \cyl\ \wh\ throat and formulate some
  further observations. We also discuss an alternative definition of a
  throat in terms of the area function $a(u)$. Comparing the consequences of
  different definitions, we arrive at an important no-go theorem for \whs\
  with flat or string asymptotic behavior at both sides of the throat: it
  turns out that, in order to make such a \wh, it is necessary to have matter
  with negative energy density. \sect 3 considers a few examples of matter
  sources of \cyl\ geometries and the corresponding \wh\ solutions. Two
  no-go theorems of more specific nature are presented there. \sect 4
  contains some concluding remarks.

\section{Cylindrical \whs: geometry and matter content}
%%%%%%%%%%%%%%%%%%%%%%%%%%%%%%%%%%%%%%%%%%%%%%%%%%%%%%%

\subsection{Basic equations}

  Let us begin with presenting the nonzero components of the Ricci tensor
  for the metric (\ref{ds-cy}) in its general form, without specifying the
  choice of the radial coordinate $u$:
\bear
    R^0_0 \eql - \e^{-2\alpha}[\gamma'' + \gamma'
				(\gamma'-\alpha'+\beta'+\xi')],
\nn
    R^1_1 \eql - \e^{-2\alpha}
		[\gamma''+\xi''+\beta''+\gamma'{}^2+\xi'{}^2+\beta'{}^2
				-\alpha'(\gamma'+\xi'+\beta')],
\nn
    R^2_2 \eql - \e^{-2\alpha}[\xi'' + \xi'
				(\gamma'-\alpha'+\beta'+\xi')],
\nn
    R^3_3 \eql - \e^{-2\alpha}[\beta'' + \beta'                  \label{Ric}
				(\gamma'-\alpha'+\beta'+\xi')],
\ear
  where the prime denotes $d/du$ and the coordinates are numbered according
  to the scheme $(0,1,2,3) = (t,u,z,\phi)$. It is also helpful to
  present the component $G^1_1$ of the Einstein tensor
  $G\mN = R\mN - \half \delta\mN R$ which does not contain any second-order
  derivatives:
\beq                                                            \label{G11}
      G^1_1 = \e^{-2\alpha}(\gamma'\xi' + \beta'\gamma' + \beta'\xi').
\eeq

  The Einstein equations are written as
\beq
  	G\mN = - \kappa T\mN, \cm  \kappa = 8\pi G,            \label{EE1}
\eeq
  where $G$ is Newton's constant of gravity, or equivalently,
\beq                                                           \label{EE2}
  	R\mN = - \kappa \tT\mN, \cm
			\tT\mN = T\mN - \Half \delta\mN T^\alpha_\alpha,
\eeq

  The above relations were written with an arbitrary $u$ coordinate. In many
  cases it is helpful to use this coordinate freedom and to choose $u$ as a
  harmonic radial coordinate, which is defined by the condition \cite{kb-flu}
\beq
	\alpha = \beta + \gamma + \xi.                       \label{harm}
\eeq
  In particular, with this choice, the expressions for $R^0_0$, $R^2_2$ and
  $R^3_3$ do not contain first-order derivatives.

\subsection{Conditions on the throat}

  Now, let us take the stress-energy tensor (SET) $T\mN$ in the most general
  form admitted by the space-time symmetry:
\beq                                                           \label{SET}
      T\mN = \diag (\rho,\ -p_r,\ -p_z,\ -p_\phi),
\eeq
  where $\rho$ is the density and $p_i$ are pressures of any physical origin
  in the respective directions.

  It is straightforward to find out how the SET components should behave on
  a \wh\ throat. At a minimum of $r(u)$, due to $\beta'=0$ and
  $\beta'' > 0$,\footnote
       {Here and henceforth we restrict ourselves for convenience to generic
	minima, at which $\beta'' > 0$. If there is a special minimum
	at which $\beta'' =0$, we still have $\beta'' >0$ in its vicinity,
	along with all consequences of this inequality. The same concerns
	minima of $a(u)$ discussed below.}
  we have $R^3_3 < 0$, and from the corresponding component of (\ref{EE2}) it
  follows
\beq                                                          \label{thr1}
    T^* := T^0_0 + T^1_1 + T^2_2 - T^3_3 = \rho - p_r - p_z + p_\phi < 0.
\eeq

  If $T^2_2 = T^3_3$, which means $p_z = p_\phi$, the condition (\ref{thr1})
  leads to $\rho - p_r < 0$, or $p_r > \rho$, which violates the {\it
  Dominant Energy Condition\/} if we assume, as usual, $\rho\geq 0$. (This is
  true, in particular, for Pascal isotropic fluids, in which all $p_i$ are
  equal to each other.) In the general case of anisotropic pressures,
  (\ref{thr1}) does not necessarily violate any of the standard energy
  conditions.

  Let us discuss what changes if we define a throat using the area function
  $a(u)$ instead of $r(u)$. We will call it an {\it a-throat\/} for clarity.

\Theorem{Definition 2}
  {In a space-time with the metric (\ref{ds-cy}), an $a$-throat is a
  cylinder $u = u_1$ where the function $a(u) = \e^{\beta+\xi}$ has a
  regular minimum.}

  By Definition 2, at $u=u_1$ we have $\beta'+\xi' =0$ and $\beta''+\xi''
  >0$. The minimum occurs in terms of {\it any\/} admissible coordinate $u$,
  in particular, in terms of the harmonic coordinate (\ref{harm}). Using it
  in \eqs (\ref{Ric}) and (\ref{EE2}), we find that the condition
  $\beta''+\xi''$ implies
\beq                                                     	\label{thr2a}
     R^2_2 + R^3_3 < 0 \ \ \then\ \ T^0_0 + T^1_1 = \rho - p_r < 0.
\eeq
  In addition, substituting $\beta'+\xi'=0$ into the Einstein equation $G^1_1
  = -\kappa T^1_1$, we find
\beq                                                            \label{thr2b}
     G^1_1 = \e^{-2\alpha}\beta'\xi' = -\e^{-2\alpha}\beta'{}^2 \leq 0
     \ \ \then \ \    -T^1_1 = p_r \leq 0.
\eeq
  Combining (\ref{thr2a}) and (\ref{thr2b}), we obtain
\beq                                                            \label{thr2}
	 \rho < p_r \leq 0\ \ \ {\rm at}\ \ u=u_1.
\eeq
  Thus at and near an $a$-throat there is necessarily a region with negative
  energy density $\rho$.

  Let us recall for comparison the \wh\ throat conditions in static spherical
  symmetry. There, the Einstein equation ${0\choose 0}-{1\choose 1}$ leads to
  the well-known inequality $\rho + p_r < 0$, which violates the {\it Null
  Energy Condition\/}. The equation $G^1_1 = -\kappa T^1_1$ leads to $p_r <0$       on the throat. Meanwhile, the density $\rho$ can have any sign.

  We conclude that both conditions (\ref{thr1}) and (\ref{thr2}) radically
  differ from their counterpart in spherical symmetry. Moreover, these two
  conditions themselves are drastically different: while (\ref{thr1}) admits
  quite usual kinds of matter (as will be seen from the examples below),
  (\ref{thr2}) definitely requires $\rho < 0$, i.e., even more exotic matter
  than in spherical symmetry.

\subsection {Asymptotic conditions; a no-go theorem}

  So far we discussed the local conditions that must hold on the throat.
  To describe a \wh\ as a global entity, it is mandatory to consider the
  geometry far from the throat, on both sides from it. We will consider a
  situation that seems the most natural, in which the \wh\ is observed as a
  stringlike source of gravity from an otherwise very weakly curved or even
  flat environment.

  So we require the existence of a spatial infinity, i.e., a value $u =
  u_\infty$ such that $r = \e^\beta \to \infty$ , where the metric is either
  flat or corresponds to the gravitational field of a cosmic string.

  First, as $u\to u_\infty$, a correct behavior of clocks and
  rulers requires $|\gamma| < \infty$ and $|\xi|<\infty$, i.e.,
\beq
    \gamma\to \const,\ \ \ \xi\to \const \ \ \
    {\rm as} \ \ u \to u_\infty.                             \label{as1}
\eeq
  Choosing proper scales along the $t$ and $z$ axes, one can turn these
  constants to zero; but if a \wh\ has two such asymptotics, this operation,
  in general, can be done only at one of them.

  Second, at large $r$ we require
\bear
    |\beta'|\e^{\beta-\alpha} \to 1 - \mu,
  \cm
    \mu = \const < 1 \cm {\rm as} \ \ \ x\to x_\infty,       \label{mu}
\ear
  so that the circumference-to-radius ratio for the circles $u=\const$,
  $z = \const$ tends to $2\pi (1-\mu)$ instead of $2\pi$ which should be the
  case if the space-time is \asflat. So the parameter $\mu$ is an angular
  defect. Under the asymptotic conditions (\ref{as1}), (\ref{mu}), $\mu>0$,
  the solution can simulate a cosmic string. A flat spatial asymptotic
  takes place if $\mu = 0$. Negative values of $\mu$ are also not {\it a
  priori\/} excluded, they correspond to an angular excess.  In what follows
  we will use the words ``{\it regular asymptotic\/}'' in the sense ``{\it
  flat or string asymptotic\/}''.

  Third, the Riemann curvature tensor, hence the Ricci tensor, should vanish
  at large $r$, and, due to the Einstein equations, all SET components must
  decay quickly enough.\footnote
       {In general, this requirement should be formulated in terms of
	Lorentz tetrad components of all tensors. However, for our diagonal
	metric (\ref{ds-cy}) these tetrad components coincide with
	coordinate components written with mixed indices, such as
	$R\MN{}_{\rho\sigma}$ for the Riemann tensor, $R\mN$ for the Ricci
	tensor and $T\mN$ for the SET. They behave as scalars at
	reparametrizations of the radial coordinate $u$, which makes a
	transition to tetrad components redundant.}

  It is easy to verify that in the coordinates (\ref{harm}) a regular
  asymptotic can only occur as $u \to \pm \infty$. Indeed, due to (\ref{as1})
  at such an asymptotic we have
\beq                                                         \label{as2}
     \alpha \sim \beta \to \infty \ \then \ |\beta'| \to \const
\eeq
  due to (\ref{mu}), and the two conditions (\ref{as2}) are compatible with
  each other only if $u \to \pm \infty$. If we deal with a \wh\ with two
  regular asymptotics, one of them occurs at $u = +\infty$, the other at
  $u=-\infty$.

  Evidently, at a regular asymptotic, both $r(u)$ and $a(u)$ tend to
  infinity. If there are two such asymptotics, both functions have minima at
  some finite $u$, i.e., there occur both a throat as a minimum of $r(u)$ and
  an $a$-throat (they do not necessarily coincide if there is no symmetry
  with respect to $u = u_1$). This leads to the following result.

\Theorem{Proposition 1}
  {In \GR, any \scyl\ \wh\ with two regular asymptotics contains a region
   where the energy density is negative.}

  This conclusion can be equivalently formulated as a no-go theorem:

\Theorem{Proposition 1a}
  {In \GR, a \scyl, twice asymptotically regular \wh\ cannot exist if the
  energy density $T^0_0$ is everywhere nonnegative.}

  It should be stressed that this conclusion holds with any of the two
  definitions of a throat.

\section{Examples of cylindrical \whs}
%%%%%%%%%%%%%%%%%%%%%%%%%%%%%%%%%%%%%%

  In what follows, we will everywhere adhere to Definition 1 of a \cy\ \wh.
  A few examples of \wh\ space-times will be presented with different matter
  sources. None of them have two regular asymptotics, even though in
  some cases the energy density is (partly) negative.

\subsection{Vacuum}

  Vacuum space-time ($T\mN =0$) cannot contain a \wh\ since the condition
  (\ref{thr1}) does not hold anywhere. Let us still consider it for
  comparison. The easiest way to solve the equations $R\mN =0$ is
  to choose $u$ as a harmonic radial coordinate, see (\ref{harm}). Then
  the equations $R\mN =0$ lead to $\beta'' = \gamma''= \xi'' =0$, so that
\beq                                                           \label{vac1}
      \gamma(u) = au + a_0,\cm \beta(u) = bu + b_0, \cm \xi(u) = cu + c_0,
\eeq
  with six integration constants, among which $a_0,\ b_0,\ c_0$ may be
  turned to zero by changing scales along the $t$ and $z$ axes and choosing
  the zero point of the $u$ coordinate. Moreover, the first-order equation
  $G^1_1 =0$ leads to a relation between the remaining constants $a,\ b,\ c$:
\beq
	ab + ac + bc =0,                                        \label{vac2}
\eeq
  hence there are two essential constants. The metric takes the form
\beq                                                            \label{vac3}
      ds^2 = \e^{2au}dt^2 - \e^{2(a+b+c)u} du^2
      					- \e^{2cu}dz^2 -\e^{2bu}d\phi^2.
\eeq
  It is the well-known Levi-Civita solution whose more usual form (see,
  e.g., \cite{bicak04})
\beq                                                           \label{vac4}
     ds^2 = \rr^{2m} dt^2 - \rr^{2m(m-1)}(d\rr^2 + dz^2)
			  	- {\cal C}\rr^{2(1-m)} d\phi^2
\eeq
  is obtained from (\ref{vac3}) using the relations and notations
\beq                                                           \label{vac5}
	\e^{(a+b)u} = k\rr, \cm k = (a+b)^{-(a+b)/c}, \
	\cm m = \frac{a}{a+b}, \cm {\cal C} = (a+b)^{2b/c}
\eeq
  valid for $a+b \ne 0$, $c\ne 0$. [Note that $a+b =0$ leads to $a=b=0$
  while $c=0$ leads to $ab=0$ due to (\ref{vac2})]. The two parameters in
  (\ref{vac4}) are: $m$ called the mass parameter and ${\cal C}$ called the
  conicity parameter.

  In the special case $m=0$ in (\ref{vac4}), or $a=c=0$ in (\ref{vac2})
  [in this case the two metrics are not related by (\ref{vac5}) but the
  result is the same] we obtain the flat metric in which $1-{\cal C}$ or
  $1-b^2$ is the angular defect.

  For our treatment it is important that even the vacuum solution is in
  general not \asflat, and only for $m=0$ or $a=c=0$ one obtains a flat
  (for $b^2=1$) or string (for $b^2 \ne 1$) asymptotic behavior. This means
  that in the general case, when $T\mN$ vanishes asymptotically and the
  metric approaches the Levi-Civita solution, the conditions (\ref{as1}) and
  (\ref{mu}) make a very strong restriction.  Recall for comparison that in
  \sph\ systems the vacuum (Schwarzschild) solution is \asflat, and the same
  is true for a wealth of non-vacuum solutions, certainly, in the absence of
  a cosmological constant.

\subsection{Scalar fields; two more no-go theorems}

  Consider a scalar field with the Lagrangian
\beq                                                            \label{L_s}
    L_s = \Half \eps\varphi^{,\alpha}\varphi_{,\alpha} - V(\varphi)
\eeq
  where $V(\varphi)$ is an arbitrary function, and $\eps=\pm 1$
  distinguishes normal ($\eps=+1$) and phantom ($\eps = -1$) scalar fields.
  For the metric (\ref{ds-cy}) and $\varphi=\varphi(u)$, the SET has the form
\bear
    T\mN \eql \eps\varphi_{,\mu}\varphi^{,\nu}                  \label{s1}
     	- \delta\mN L_s
	= \Half \eps {\varphi'}^2 \e^{-2\alpha}\diag (1, -1, 1, 1)
			      + V(\varphi)\delta\mN,
\ear
  so that
\beq                                                            \label{s2}
  	T_0^0 = T_2^2 = T_3^3.
\eeq
  Therefore, in the coordinates (\ref{harm}), three second-order equations
  (\ref{EE2}) combine to give
\beq
    \beta''=\gamma''= \xi'' = \thd\alpha''                      \label{s3}
\eeq
  where the last equality is due to (\ref{harm}), whence
\beq
		\xi = \thd(\alpha-Au),
	\cm
	     \gamma = \thd(\alpha-Bu),
	\cm          				                \label{s4}
	      \beta = \thd(\alpha + Au + Bu),
\eeq
  where $A$ and $B$ are integration constants and two more constants are
  ruled out by moving the origin of $u$ and rescaling the $z$ axis. The
  remaining unknowns $\alpha$ and $\varphi$ obey the equations [the scalar
  field equation and combinations of (\ref{EE1})]
\bearr
    \alpha'' + 3\kappa V(\varphi) \e^{2\alpha} = 0,     	\label{s5}
\yyy
    \eps\varphi'' - (dV/d\varphi) \e^{2\alpha} = 0.             \label{s6}
\yyy
    \alpha'{}^2 - N^2 = \fract{3}{2}\eps\kappa \varphi'{}^2     \label{s7}
					    -3\kappa V\e^{2\alpha},
	\cm    	  N^2 := \thd(A^2 +AB + B^2),
\ear
  where (\ref{s7}), following from the ${1\choose 1}$ component of
  (\ref{EE1}), is a first integral of (\ref{s5}) and (\ref{s6}).

  For a SET satisfying (\ref{s2}), the condition (\ref{thr1}) leads to
  $V < 0$ for both normal and phantom fields. Thus \whs\ solutions are not
  excluded but require a negative potential, at least on and near the throat.

  Suppose now that there is a {\it regular spatial asymptotic}.
  Without loss of generality it can be placed at $u\to +\infty$, where due to
  (\ref{as2}) $\alpha \approx \beta$ and due to (\ref{s4}) we have
\[
      A = B = N >0, \cm	\alpha \approx \beta \approx Nu.
\]
  Another regular asymptotic might occur at $u \to -\infty$; however, since
  the relation for the integration constants $A=B=N$ still holds, if we
  assume that $\gamma$ and $\xi$ are finite there, we arrive again at
  $\alpha \sim \beta \sim Nu$, but now it means that $\beta\to -\infty$,
  that is, an axis (which can in principle be regular); another regular
  spatial infinity cannot exist. We arrive at the following result
  (see also \cite{br-sh01}):

\Theorem{Proposition 2}
   {\it Static, \cyl\ wormholes with two regular asymptotics do not exist in
    \GR\ with matter whose SET satisfies \eq (\ref{s2}).}

  If we deny the asymptotic regularity condition but require symmetry of the
  \wh\ with respect to its throat, then at such a throat ($u=u_0$)
  $\beta' = \gamma' = \xi' =0$, and \eqs (\ref{EE1}) may be combined to give
\beq
	\beta'' = -\e^{2\alpha}(T^0_0 + T^2_2 - T^3_3)        \label{minB}
\eeq
  at $u=u_0$. Since it is a minimum of $\beta$, we have there $\beta'' > 0$.
  Assuming $T^2_2 = T^3_3$ (which is true for scalar fields), \eq
  (\ref{minB}) means $T_0^0 < 0$. The result is (see also \cite{br-sh01}):

\Theorem{Proposition 3}
    {\it A static, \cyl\ wormhole, symmetric with respect to its throat,
    cannot exist in \GR\ with matter whose SET satisfies the conditions
    $T^2_2 = T^3_3$ and $T^0_0 \geq 0$.}

  Propositions 2 and 3 not only apply to scalar fields with the Lagrangian
  (\ref{L_s}) but to any matter whose SET satisfies the corresponding
  condition, for instance, scalar fields with more general Lagrangians
  like $F(\phi, X)$, $X = g\MN\phi_{,\mu}\phi_{,\nu}$ frequently used to
  model dark energy (k-essence, generalized Chaplygin gas models etc.)
  as well as spinor fields to be briefly discussed further in this section.
  Proposition 3 also applies to any Pascal (not necessarily perfect) fluids
  with isotropic pressure.

\subsection{Scalar fields: some \wh\ solutions}

  Consider a solution to \eqs (\ref{s5})--(\ref{s7}) in a special case of
  negative potential, putting
\beq
     3\kappa V = 3\Lambda = -\lambda^2 < 0,\cm \lambda > 0,
\eeq
  where $\Lambda <0$ is a cosmological constant. Thus we are dealing with a
  self-gravitating massless scalar field in the presence of a cosmological
  constant. We can expect \wh\ solutions but certainly without regular
  asymptotics, not only due to Proposition 2 but simply because the
  curvature must be nonzero at an asymptotic if $\Lambda \ne 0$.

  From \eqs (\ref{s5}) and (\ref{s6}) we get
\beq                                                              \label{s8}
	\varphi' = C = \const, \cm  \alpha'' = \lambda^2 \e^{2\alpha}.
\eeq
  The latter is a Liouville equation whose solution may be written as
\beq
    \e^{-\alpha} = \vars{ (\lambda/h) \sinh [h(u-u_1)], &  h>0,\\
			   \lambda (u-u_1), & h = 0,\\
			  (\lambda/h) \sin [h(u-u_1)], &   h<0,
			}                                        \label{s9}
\eeq
  where $h$ and $u_1$ are integration constants. \eq (\ref{s7}) leads to a
  relation among the constants:
\beq
    h^2 \sign h = N^2 + \frac{3}{2}\eps \kappa C^2.              \label{s10}
\eeq
  This, together with (\ref{s4}), completes the solution.

  Within this general solution, let us now single out \wh\ solutions, i.e.,
  those in which $\beta(u)$ tends to infinity at both ends of the range of
  $u$. For convenience and without loss of generality we put $u_1 = 0$ and
  assume $u > 0$. It is then easy to see that in the limit $u\to 0$ we have
  $\beta\to \infty$ for any values of $h$ since $\alpha \approx -\ln u \to
  \infty$. Let us look when $\beta\to\infty$ at large $u$.

  If $h >0$ [by (\ref{s10}), it is always the case if there is a normal
  scalar field, $\eps = +1$, $C\ne 0$, but is also possible with a phantom
  scalar, $\eps=-1$, $C \ne 0$, and in the absence of a scalar, $C=0$],
  then $\alpha \approx -hu$ as $u\to\infty$, hence $3\beta \approx (A+B-h)u$,
  and we obtain a \wh\ if $A + B > h$.

  The value $h=0$ may appear without a scalar field ($C=0$, a pure vacuum
  solution with negative $\Lambda$), it corresponds to $A = B = 0$, and
  one can verify that this is a \cyl\ version of the anti-de Sitter metric
  which is not \wh. However, by (\ref{s10}), we may have $h=0$ with nonzero
  $A$ or $B$ (or both) if there is a phantom scalar ($\eps=-1$, $C\ne 0$).
  Then, as $u \to \infty$, $\alpha \sim - \ln u \to -\infty$, but its
  contribution is inessential in the expressions (\ref{s4}), and we have
  $\beta \to \infty$, hence a \wh, if $A + B >0$.

  Lastly, if $h < 0$, which can only happen in the presence of a phantom
  field, the other end of $u$ range is $u \to \pi/|h|$, and it is quite
  similar to $u=0$. Thus {\it all\/} solutions with $h < 0$ are \wh.

  So, we have a large family of \wh\ solutions with a negative cosmological
  constant, with or without massless scalar fields, both normal and phantom.
  However, none of these solution are asymptotically regular.

  Among these solutions, there is a symmetric subfamily: it corresponds to
  $\eps =-1$, $C \ne 0$, $A = B = 0$, $h < 0$. In accord with Proposition 2,
  it has a negative energy density, $T^0_0 = -\half \phi'{}^2
  \e^{-2\alpha} + \Lambda/\kappa$. In terms of the Gaussian coordinate
  $l = (1/\lambda)\log \tan (hu/2)$ ($l \in \R$ is a length in the radial
  direction), the metric in this case can be written in the simple-looking
  form
\beq
    ds^2 = -dl^2
         + \biggl(\frac{|h|}{\lambda}\cosh\frac{l}{\lambda}\biggr)^{2/3}
	     (dt^2 - dz^2 - d\varphi^2).
\eeq

\subsection{Spinor fields: general considerations}

  Spinor fields of sufficiently general nature in \scyl\ configurations have
  been considered in Ref.\,\cite{spin04}, and we here follow this paper.
  The Lagrangian
\bear
     L_{\spn} = \frac{i}{2}\big(\opsi \gamma^{\mu} \nabla_{\mu} \psi -
            (\nabla_{\mu} \opsi) \gamma^{\mu} \psi\ \big)
                - m \opsi \psi - F(S),             	    \label{L_sp}
\ear
  where $F(S)$ is an arbitrary function of the invariant $S = \opsi \psi$,
  describes as a special case the Dirac spinor field $\psi$ of arbitrary
  mass $m$ as well as a general class of nonlinearities. The equations for
  $\psi$ and $\opsi$ and the spinor field SET are \cite {zheln}
\bearr                                                      \label{e-sp1}
     i\gamma^{\mu} \nabla_{\mu}\psi - m\psi - \d F/\d\opsi = 0,
\yyy                                                        \label{e-sp2}
     i\nabla_{\mu}\opsi\gamma^\mu + m\opsi + \d F/\d\psi = 0,
\yyy  \nq                                                   \label{SET-sp}
     T_{\mu\ \spn}^{\rho} =
        \frac{i}{4} g^{\rho\nu}(\opsi\gamma_\mu \nabla_\nu \psi
            + \opsi\gamma_\nu \nabla_\mu \psi
                - \nabla_\mu \opsi\gamma_\nu \psi
       -\nabla_\nu \opsi\gamma_\mu \psi) - \delta^\rho_\mu L_\spn.
\ear
  where $\nabla_\mu \psi$ is the covariant derivative of the spinor field
\bear
    \nabla_{\mu} \psi = \d\psi/\d x^\mu - \Gamma_{\mu} \psi,  \label{d-psi}
\ear
  $\Gamma_{\mu}(x) $ being the Fock-Ivanenko spinor affine connection
  matrices. The $\gamma^\mu$ matrices are related to the flat-space Dirac
  matrices $\tgamma^{\mu}$ by  $\gamma_\mu (u) = e^a_\mu\, \tgamma_{a}$,
  where $e^a_\mu $ are the tetrad vectors, so that  $g\mn
  = e^a_\mu \,\e^b_\nu\, \eta_{ab}$, where $\eta_{ab} = \diag(1,-1,-1,-1)$.

  In our \scyl\ case, with the metric (\ref{ds-cy}) and $\psi = \psi(u)$,
  we have \cite{spin04} $L_{\spn} = S F_S - F(S)$, $F_S := dF/dS$. The SET
  components of the spinor field are
\bearr
       T_0^0 = T_2^2 = T_3^3 = F(S) - SF_S,                  \label{T-sp}
\\ \lal
       T_1^1 = \frac{i}{2}(\opsi\gamma^1 \nabla_1 \psi -
      \nabla_1 \opsi\gamma^1 \psi) + SF_S - F(S),          \label{T11-sp}
\ear
  so that \eq (\ref{s2}) holds, along with its consequences such as the
  expressions (\ref{s4}) for the metric functions and Propositions 2 and 3
  which restrict the possible \wh\ existence.

  \eq (\ref{e-sp1}) may be written in the form \cite{spin04}
\bear
    i\e^{-\alpha}\tgamma^{1}(\d_u + \half \alpha')\psi
          - m\psi - F_S \psi=0.                             \label{sp8}
\ear
  Combining it with its conjugate, we arrive at the equation
  $S' + \alpha' S =0$, giving
\bear
      S(u) = c_0 \e^{-\alpha(u)},\qquad c_0 = \const.       \label{sp9}
\ear
  Then $F$ and $F_S$ are expressed in terms of $\e^{-\alpha(u)}$. Moreover,
  \eq (\ref{sp8}) and its conjugate allow one to re-express $T_1^1$ as
\bear
        T_1^1 = m S + F(S) =: M(S).                         \label{sp10}
\ear

  The only remaining Einstein equation to be solved is
\bear                                       		    \label{sp11}
     {\alpha'}^2 - N^2 = - 3\kappa \e^{2\alpha} M(S), \cm
	    		N^2 := \thd (A^2 + A B + B^2),
\ear
  Since by (\ref{sp9}) $\alpha'= - S'/S$, \eq (\ref{sp11}) is
  rewritten as
\bear                                           	    \label{sp12}
     \biggl(\frac{dS}{d u}\biggr)^2 = N^2 S^2
            			- \frac{3\kappa c_0^2} {M(S)},
\ear
  which is easily solved by quadratures. Thus, given $F(S)$, the Einstein
  equations are solved in a general form even without entirely integrating
  the nonlinear spinor equations \cite{spin04}.

  With (\ref{T-sp}) and (\ref{sp10}), the condition (\ref{thr1}) implies
\beq
	2M - SM_S < 0                                     \label{sp13}
\eeq
  at a \wh\ throat. This condition is similar to the
  $V < 0$ condition for scalar fields.

\subsection{Spinor field: example of a \wh\ solution}

  A simple example of a \wh\ solution can be obtained using the inverse
  problem method: choosing the form of $\alpha(u)$, we easily find both
  $S(u)$ and $M(u)$, hence $M(S)$. So, let us put
\beq                                                          \label{sp14}
       \e^{\alpha} = A_0 \cosh ku, \cm A_0,\ k = \const >0,
\eeq
  to obtain, according to (\ref{sp9}) and (\ref{sp11}),
\bearr                                                        \label{sp15}
	S(u) = c_0/(A_0 \cosh ku),
\yyy
    3\kappa A_0^2 M(S) = \frac{N^2-k^2}{\cosh^2 ku} + \frac{k^2}{\cosh^4 ku}
	 = (N^2 -k^2)
	     \biggl(\frac{A_0}{c_0}\biggr)^2 S^2 +
	    	k^2 \biggl(\frac{A_0}{c_0}\biggr)^4 S^4.      \label{sp16}
\ear
  The solution as a whole is determined by \eqs (\ref{s4}), (\ref{sp14}),
  and (\ref{sp15}). It is regular at all $u \in \R$ and, since the
  asymptotics of $\beta(u)$ are
\bear
	 \beta \approx \vars{  (k+A+B) u, &  u\to +\infty,\\
			       (k-A-B)|u|, & u \to -\infty,   }
\ear
  it describes a \wh\ if $|A+B| < k$. The \wh\ is symmetric if $A=B=0$,
  which corresponds to a nonlinear spinor field with
\[
   	M(S) = -\frac{k^2 S^2}{3\kappa c_0^2}
		\biggl[ 1 - \biggl(\frac{A_0 S}{c_0}\biggr)^2\biggr].
\]
  Its metric can be written in terms of the Gaussian coordinate
  $l = (A_0/k) \sinh ku \in \R$) as
\beq
    ds^2 = - dl^2 + (A_0^2 + k^2 l^2)^{1/3} (dt^2 -dz^2 -d\phi^2).
\eeq
  It is easy to verify that, as in the scalar case, the density $T^0_0$ is
  negative near the throat $u=0$, $l=0$.

\subsection{Einstein-Maxwell fields and nonlinear electrodynamics (NED)}

  Electromagnetic fields $F\mn$, compatible with the geometry (\ref{ds-cy}),
  can have three different directions:
%\begin{description}  \itemsep 0pt

\noindent
Radial (R) fields: electric, $F_{01} (u)$ ($E^2 =F_{01}F^{10}$), and
      magnetic, $F_{23} (u)$ ($B^2 = F_{23}F^{23}$).

\noindent
Azimuthal (A) fields: electric, $F_{03} (u)$ ($E^2 =F_{03}F^{30}$),
      and magnetic, $F_{12}(u)$ ($B^2 = F_{12} F^{12}$).

\noindent
Longitudinal (L)] fields: electric, $F_{02}(u)$  ($E^2
      =F_{02}F^{20}$), and magnetic, $F_{13}(u)$ ($B^2 = F_{13}F^{13}$).

  Here $E$ and $B$ are the absolute values of the electric field strength
  and magnetic induction, respectively. Self-gravitating \scyl\
  configurations of such electromagnetic fields have been considered in the
  framework of the Einstein-Maxwell theory in \cite{br79} and in the
  Einstein-NED theory with gauge-invariant NED Lagrangians of the form
\beq
  	L_e = -\Phi(F)/(16\pi), \cm F := F\MN F\mn              \label{L_e}
\eeq
  in \cite{NED-cy}. The Maxwell Lagrangian is recovered by putting
  $\Phi(F) \equiv F$. The Lagrangian (\ref{L_e}) leads to the SET
\def\fpi{\frac{1}{16\pi}}
\beq                                                          \label{SET-e}
	T\mN = \fpi [-4F^{\nu\alpha}F_{\mu\alpha}\Phi_F + \delta\mN \Phi],
\eeq
  where $\Phi_F = d\Phi/dF$. For a radial electromagnetic field in the
  geometry (\ref{ds-cy}) this gives
\bearr
     T_0^0 = T_1^1 = \fpi (4E^2 \Phi_F + \Phi),        		\label{01R}
\\ \lal
     T_2^2 = T_3^3 = \fpi (-4B^2 \Phi_F + \Phi).       		\label{23R}
\ear
  Similar relations for a longitudinal field are obtained by replacing
  $1 \leftrightarrow 2$ in the indices and for an azimuthal field by
  replacing $1 \leftrightarrow 3$.

  Now, let us find out which kind of electromagnetic fields is suitable
  for obtaining \whs. It is easy to verify that the expression (\ref{thr1}),
  which should be negative at and near a \wh\ throat, has the form
\bearr                                                           \label{e5}
    T^* = (4E^2 \Phi_F + \Phi)/(8\pi) \ \ \mbox{(R- and L- fields)},
\yyy                                                             \label{e6}
    T^* = (-4B^2\Phi_F + \Phi)/(8\pi) \ \ \mbox{(A- fields)}.
\ear

  In Maxwell electrodynamics this gives $T^* = (E^2+B^2)/(4\pi)$ for L- and
  R-fields and $T^* = -(E^2+B^2)/(4\pi)$ for A fields. Thus \whs\ are only
  possible with azimuthal fields. Indeed, the corresponding exact solution
  to the Einstein-Maxwell equations has the form \cite{br79}
\bearr                                                           \label{e7}
     ds^2 = \frac{\cosh^2 (hu)}{Kh^2} \bigl[\e^{2au}d t^2
     	  	- \e^{2(a+b)u}du^2
     - \e^{2bu} d \varphi^2 \bigr] - \frac{K h^2}{\cosh^2 (hu)} dz^2,
\ear
  where $K = [G(i_e^2 + i_m^2)]^{-1} $, $h^2 = ab$, $a, b = \const$,
  $a > 0$, $b > 0$, and the electromagnetic field is given by
\beq
      F_{03} = i_m = \const; \quad\                             \label{e8}
      F^{12} = i_e \e^{-2\alpha}, \ \ i_e = \const,
\eeq
  where $i_e$ and $i_m$ are the effective currents of electric and magnetic
  charges along the $z$ axis, respectively. This solution is written under
  the coordinate condition (\ref{harm}), and evidently, $r = \e^\beta \sim
  \e^{(h\pm b) |u|}$ as $u \to \pm \infty$. Thus it is a \wh\ solution if $b
  < h$. It is easy to see that such a \wh\ is neither symmetric nor
  asymptotically regular.

  Let us now turn to NED, see \eq (\ref{L_e}). It is clear that $\Phi(F)$ may
  be chosen so that the expression (\ref{e5}) for R- and L-fields will be
  negative at some $F$, but this will simultaneously mean that the energy
  density $T^0_0$ will be negative. Such cases are yet to be studied.
  On the other hand, with A-fields it is easy to obtain \wh\ solutions like
  (\ref{e7}), (\ref{e8}), and moreover one can show that no such solutions
  have two regular asymptotics. Indeed, since in this case $T^0_0 = T^3_3$,
  the corresponding component of the Einstein equations in the coordinates
  (\ref{harm}) gives $\beta'' = \gamma''$, whence $\beta = \gamma + bu + b_0$
  with $b,\ b_0 = \const$.  A regular asymptotic at $u = \infty$ requires
  $\gamma \to \const$ and $b >0$ whereas a regular asymptotic at $u = \infty$
  requires $\gamma \to \const$ and $b <0$. Thus a regular asymptotic can
  appear only at one ``end'', and, moreover, it can be only achieved at the
  expense of a non-Maxwell behavior of $\Phi(F)$ at small $F$ \cite{NED-cy}.

\section{Concluding remarks}

  We have seen that the existence conditions for \cyl\ \whs\ can be
  satisfied without violating the weak or null energy conditions near the
  throat. We have presented a number of explicit examples of \wh\ solutions
  with non-phantom sources.

  However, as is always the case when dealing with \cyl\ systems, it
  is rather hard to obtain solutions with regular (i.e., flat or string)
  asymptotics: indeed, even the Levi-Civita vacuum solution has such an
  asymptotic only in a special case. Of course, such asymptotic behaviors
  are necessary if we wish to describe a \wh\ in a flat or weakly curved
  background universe. We have proved that if one wishes to have this
  behavior at both sides of the throat, it is necessary to invoke matter with
  negative energy density.

  This problem is still more important if we try to apply \cyl\ solutions
  as an approximate description of toroidal systems, e.g., like those
  discussed by Gonzalez-Diaz \cite{gdiaz}. This approximation
  must work well if a torus containing matter and significant curvature is
  thin, like a circular string, i.e., its larger radius is much greater than
  the smaller radius. In this case, small segments along such a ``string''
  are approximately cylindrically symmetric. But this means that
  sufficiently far from the thin ring, in any direction (to or from the
  center of the ring or in any other), the space-time should be almost flat.

  Thus, in future studies, it is desirable either to find asymptotically
  regular \cyl\ geometries with more or less realistic material sources or
  to consider configurations consisting of at least three layers: one
  responsible for the \wh\ throat and its neighborhood and two others (on
  each side of the throat) providing regular asymptotics. One can note that
  the required negative energy densities may appear due to quantum effects,
  such as vacuum polarization and the Casimir effect.

  A challenging problem is to clear up a relationship between the throat
  topology and the energy conditions. A point of interest is that in
  \cyl\ space-times there can be at least two different definitions of a \wh\
  throat:  the one we have been using, in terms of the radius $r(u)$, and the
  alternative one, in terms of the area function $a(r)$; moreover, they lead
  to different \wh\ existence conditions in terms of the SET, but both of
  them differ from the conditions for \sph\ \whs\ or those with a
  spherical topology of throats. Thus the topological issue is probably
  crucial for formulating the general properties of matter sources for \wh\
  geometries. And it is yet to be ascertained what are the similar conditions
  at throats of toroidal and more complex topologies. It seems plausible that
  a toroidal throat should behave like a cylindrical one since the condition
  sought for should be of local nature, but a rigorous proof is so far
  lacking.

  There are quite a number of results indicating \wh\ properties of many
  stringlike configurations. One can mention, in particular, Clement's ``flat
  \whs'' obtained from properly moving straight cosmic strings, the \wh\
  nature of Kerr and Kerr-Newman space-times with large charges and/or
  angular momenta, where naked ring singularities have certain string
  properties (see \cite{clem99, bur} and references therein), and their static
  counterparts \cite{zipoy,kb-ringwh}.\footnote
  	{There is a fundamental difference between ``ring \whs'' described
	by Kerr, Zipoy and similar metrics and the cylindrical and toroidal
	\whs\ discussed here: in the first case, one gets from one
	``universe'' to another by threading the ring, i.e., crossing the
	disc that subtends it, while in the second case, to traverse the
	\wh, it is necessary to approach the ``string'' itself.}
  Rotating \cyl\ configurations also tend to show \wh\ properties
  \cite{krechet}.
  There are also toroidal wormhole solutions in anti-de Sitter space-times
  built using the cut and paste technique \cite{lemoslobo}. So one can be
  rather optimistic about the existence of realistic cylindrical or toroidal
  \wh\ models.

\subsection*{Acknowledgments}
%%\begin{acknowledgments}
  We thank Sean Hayward and Hideki Maeda for helpful discussions.
  This work was partially funded by Funda\c c\~ao para a
  Ci\^encia e Tecnologia (FCT) --- Portugal, through project
  PPCDT/FIS/57552/2004. KB was funded in part by NPK MU grant at Peoples'
  Friendship University of Russia. KB is grateful to colleagues from CENTRA,
  where part of this work was done, for kind hospitality.
%%\end{acknowledgments}

\end{document}